\documentclass[twocolumn,floatfix,superscriptaddress,showpacs]{revtex4}
\usepackage{graphicx}
\usepackage{amsmath}
\usepackage{bm}
\begin{document}

\title{Quantum-limited shot noise in graphene}
\author{J. Tworzyd{\l}o}
\affiliation{Institute of Theoretical Physics, Warsaw University, Ho\.{z}a 69, 00--681 Warsaw, Poland}
\author{B. Trauzettel}
\affiliation{Instituut-Lorentz, Universiteit Leiden, P.O. Box 9506, 2300 RA Leiden, The Netherlands}
\author{M. Titov}
\affiliation{Department of Physics, Konstanz University, D--78457 Konstanz, Germany}
\author{A. Rycerz\footnote{On leave from: Marian Smoluchowski Institute of Physics, Jagiellonian University, Krak\'{o}w, Poland.}}
\affiliation{Instituut-Lorentz, Universiteit Leiden, P.O. Box 9506, 2300 RA Leiden, The Netherlands}
\author{C. W. J. Beenakker}
\affiliation{Instituut-Lorentz, Universiteit Leiden, P.O. Box 9506, 2300 RA Leiden, The Netherlands}
\date{March 2006}
\begin{abstract}
We calculate the mode-dependent transmission probability of massless Dirac fermions through an ideal strip of graphene (length $L$, width $W$, no impurities or defects), to obtain the conductance and shot noise as a function of Fermi energy. We find that the minimum conductivity of order $e^{2}/h$ at the Dirac point (when the electron and hole excitations are degenerate) is associated with a maximum of the Fano factor (the ratio of noise power and mean current). For short and wide graphene strips the Fano factor at the Dirac point equals $1/3$, three times smaller than for a Poisson process. This is the same value as for a disordered metal, which is remarkable since the classical dynamics of the Dirac fermions is ballistic.
\end{abstract}
\pacs{73.50.Td, 73.23.-b, 73.23.Ad, 73.63.-b}
\maketitle

Two recent experiments \cite{Nov05,Zha05} have discovered
that the conductivity of graphene (a single atomic layer of
carbon) tends to a minimum value of the order of the quantum unit
${e^{2}/h}$ when the concentration of charge carriers tends
to zero. This quantum-limited conductivity is an intrinsic
property of two-dimensional Dirac fermions (massless excitations
governed by a relativistic wave equation), which persists in an
ideal crystal without any impurities or lattice defects \cite{Lud94,Zie98,Per05,Kat05}. In
the absence of impurity scattering, and at zero temperature, one
might expect the electrical current to be noiseless. In contrast,
we show that the minimum in the conductivity is associated with a maximum in the Fano factor
(the ratio of noise power and mean current).
The Fano factor at zero carrier concentration takes on the universal value ${1/3}$ 
for a short and wide graphene strip. This is three times
smaller than the Poissonian noise in a tunnel junction and
identical to the value in a disordered metal \cite{Bee92,Nag92} --- even though the 
classical dynamics in the graphene strip is ballistic.

Shot noise measurements have proven to be a valuable diagnostic
tool in carbon nanotubes, which can be thought of as rolled-up
sheets of graphene. Very low shot noise in well-contacted bundles
of single-wall nanotubes is an indication of nearly ballistic
one-dimensional transport \cite{Roc02}. Super-Poissonian noise has
been found in a quantum dot formed out of a single-wall nanotube,
and explained in terms of inelastic tunneling in this
zero-dimensional system \cite{Ona06}. Our prediction of
sub-Poissonian shot noise in two-dimensional graphene is another
manifestation of the importance of dimensionality for quantum
transport.

\begin{figure}
\centerline{\includegraphics[width=0.9\linewidth]{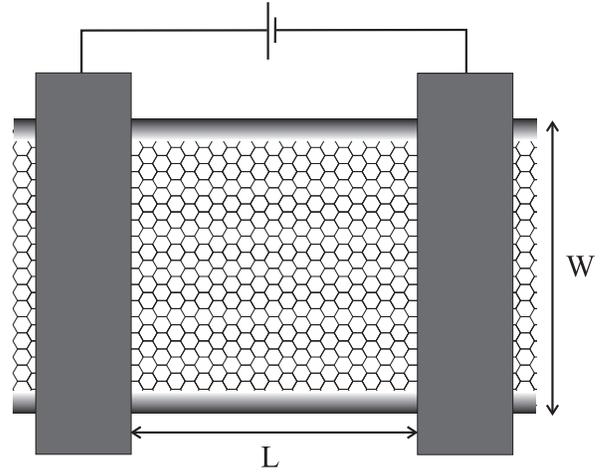}}
\caption{\label{setup} Schematic of a strip of graphene of width $W$, contacted by two electrodes (black rectangles) at a distance $L$. A voltage source drives a current through the strip. A separate gate electrode (not shown) allows the carrier concentration in the strip to be tuned around the neutrality point.}
\end{figure}

Our analysis of the shot noise was inspired by an insightful recent paper of Katsnelson \cite{Kat05}, who used the Landauer transmission formula to obtain the quantum-limited conductivity. Following the same approach, we calculate the transmission probabilities of Dirac particles through a strip of graphene in the geometry of Fig.\ \ref{setup}. (An earlier study of the same geometry counted the number of propagating modes, without determining their transmission probabilities \cite{Per06}.) The result depends on the aspect ratio $W/L$ of the strip and also on microscopic details of the upper and lower edge. For short and wide strips ($W/L\gg 1$) these microscopic details become insignificant. For that reason we first discuss the simplest case of an edge which is smooth on the scale of the lattice spacing. This corresponds to confinement of the carriers by lattice straining. The opposite case of an abrupt edge (corresponding to confinement by etching) is considered later on. 

The band structure of graphene has two valleys, which are decoupled in the case of a smooth edge. In a given valley the excitations have a two-component envelope wave function $\Psi=(\Psi_{1},\Psi_{2})$, varying on scales large compared to the lattice spacing. This continuum description of the electronic states in graphene has been found to be quite accurate \cite{Bre06}, and we will test it later on by comparing with a numerical solution of the scattering in a tight-binding model. The two components of $\Psi$ refer to the two sublattices in the two-dimensional honeycomb lattice of carbon atoms. (The additional spin degeneracy of the excitations does not play a role here.) The wave equation for $\Psi$ is the Dirac equation,
\begin{equation}
\label{Dirac}
\bigl[vp_{x}\sigma_{x}+vp_{y}\sigma_{y}+v^{2}M(y)\sigma_{z}+eV(x)\bigr]\Psi(\bm{r}) =\varepsilon \Psi(\bm{r}),
\end{equation}
with $v$ the velocity of the massless excitations of charge $e$ and energy $\varepsilon$, $\bm{p}=-i\hbar\partial/\partial\bm{r}$ the momentum operator, $\bm{r}=(x,y)$ the position, and $\sigma_{i}$ a Pauli matrix. We choose the zero of energy such that the Fermi level is at $\varepsilon=0$.

The mass term $M(y)$ is zero in the interior of the strip and rises to $\infty$ at the edges $y=0$ and $y=W$, thereby confining the particles. As shown by Berry and Mondragon \cite{Ber87}, infinite mass confinement corresponds to the boundary condition
\begin{equation}
\label{bc}
\Psi_1\big|_{y=0}=\Psi_2\big|_{y=0}, \;\;
\Psi_1\big|_{y=W}=-\Psi_2\big|_{y=W}.
\end{equation}
As a result of this boundary condition, the transversal momenta are quantized as
\begin{equation}
\label{qc}
q_n=\frac{1}{W}\pi\left(n+\tfrac{1}{2}\right),\;\; n=0,1,2, \dots ,
\end{equation}
with $n$ labeling the modes. The quantization condition for Dirac particles confined by an infinite mass differs from the one for normal electrons confined by an infinite potential by the offset of $1/2$, originating from the $\pi$ phase shift in the boundary condition  (\ref{bc}).

The electrostatic potential $V(x)=V_{\rm gate}$ for $0<x<L$, varied by a gate voltage, determines the concentration of the carriers in the strip. The value $V_{\rm gate}=0$ corresponds to charge neutrality, being the point where electron and hole excitations are degenerate (known as the Dirac point). We model the electrodes by taking a large value $V(x)=V_{\infty}$ in the leads $x<0$ and $x>L$. (The parameter $V_{\infty}$ will drop out of the results, if $|V_{\infty}|\gg |V_{\rm gate}|$.)

We calculate the transmission probabilities at the Fermi level by matching modes at $x=0$ and $x=L$. The matching condition for the Dirac equation is the continuity of the two components of $\Psi$. This ensures the conservation of the local current density $\bm{j}(\bm{r}) =ev \Psi^\dagger(\bm{r}) \cdot \bm{\sigma}\cdot\Psi(\bm{r})$, with $\bm{\sigma} = (\sigma_x, \sigma_y)$, without requiring continuity of derivatives (as needed for the Schr\"{o}dinger equation). There is a separate transmission probability $T_{n}$ for each of the $N$ propagating modes in the leads, because the matching condition does not mix the modes. (The integer $N\gg 1$ is given by $N={\rm Int}\,\bigl(k_{\infty}W/\pi+\tfrac{1}{2}\bigr)$, with $e|V_{\infty}|=\hbar vk_{\infty}$.) 

Details of the calculation are given in the Appendix. At the Dirac point $V_{\rm gate} = 0$ the transmission probability reads
\begin{eqnarray} 
\label{tn_hf}
T_{n}&=&\frac{1}{\cosh^2 Lq_n +(q_n/k_{\infty})^2 \sinh^2 Lq_n }\nonumber\\
&\rightarrow&\frac{1}{\cosh^2[ \pi (n+1/2)L/W]}\;\;{\rm for}\;\;N\gg W/L.
\end{eqnarray}
The formula (\ref{tn_hf}) is essentially different from the textbook formula \cite{note1} for the transmission probability of nonrelativistic electrons through a potential barrier, which vanishes in the limit $N\rightarrow\infty$ at zero energy (relative to the top of the barrier). 

The finite transmission probability at the Dirac point tends to the ballistic limit $T_{n}\rightarrow 1$ with increasing $|V_{\rm gate}|$. For $N\rightarrow \infty$ we find the expression
\begin{equation}
T_{n}=\left|\frac{2\kappa^{2}-2(q_{n}-k_{n})^{2}}{e^{k_{n} L}(q_{n}-k_{n}+i\kappa)^{2}+e^{-k_{n} L}(q_{n}-k_{n}-i\kappa)^{2}}\right|^{2},\label{Tna}
\end{equation}
with $\kappa=e|V_{\rm gate}|/\hbar v$ and $k_{n}=\sqrt{q_{n}^{2}-\kappa^{2}}$ for $q_{n}>\kappa$ or $k_{n}=i\sqrt{\kappa^{2}-q_{n}^{2}}$ for $q_{n}<\kappa$.

\begin{figure}
\centerline{\includegraphics[width=0.9\linewidth]{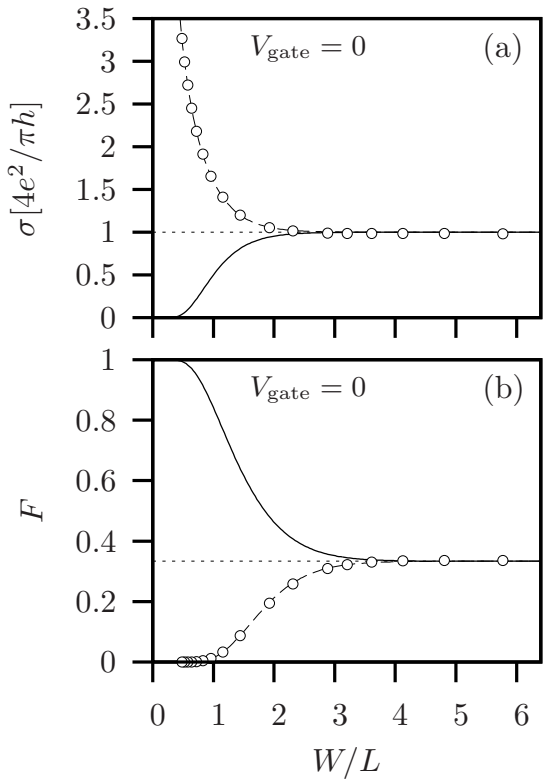}}
\caption{\label{Sigma0}
Conductivity $\sigma\equiv G\times L/W$ and Fano factor $F$ at the Dirac point ($V_{\rm gate}=0$), as a function of the aspect ratio of the graphene strip. The curves are calculated from Eq.\ (\ref{G}) in the limit $N\rightarrow\infty$, for two different boundary conditions: smooth edge [solid curves, using Eq.\ (\ref{tn_hf})] and ``metallic armchair'' edge [dashed curves, using Eq.\ (\ref{tn_armchair})]. The limit $W/L\rightarrow\infty$ (dotted lines) is given by Eq.\ (\ref{limits}), regardless of the boundary condition. The data points for the armchair edge are the result of a numerical solution of the tight-binding model on a hexagonal lattice.}
\end{figure}

\begin{figure}
\centerline{\includegraphics[width=0.9\linewidth]{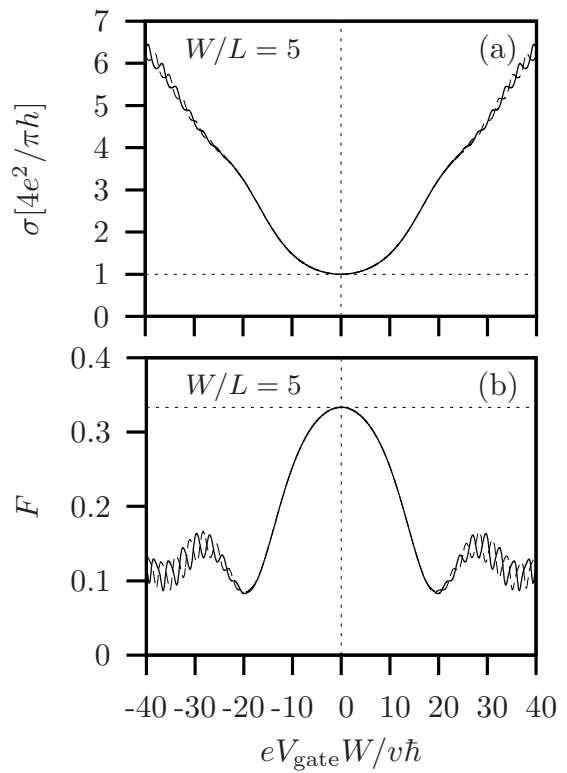}}
\caption{\label{SigmaGate}
Gate voltage dependence of the conductivity and Fano factor for a fixed aspect ratio. The conductivity minimum at the Dirac point corresponds to maximal Fano factor. The curves are calculated from Eq.\ (\ref{Tna}) for the case of a smooth edge (solid curves, taking $q_{n}=(n+1/2)\pi/W$) or metallic armchair edge (dashed curves, taking $q_{n}=n\pi/W$). The oscillations signal the appearance of propagating modes in the graphene strip with increasing gate voltage.}
\end{figure}

The conductance $G$ and Fano factor $F$ follow by summing over the modes,
\begin{equation}
\label{G}
G=g_{0}\sum_{n=0}^{N-1} T_{n} ,\;\; F=\frac{\sum_{n=0}^{N-1}T_{n}(1-T_{n})}{\sum_{n=0}^{N-1}T_{n}},
\end{equation}
with $g_{0}=4e^{2}/h$. (The factor $4$ accounts for the spin and valley degeneracy.) The dependence of the conductivity $\sigma\equiv G\times L/W$  and the Fano factor at $V_{\rm gate}=0$ on the aspect ratio $W/L$ is plotted in Fig.\ \ref{Sigma0} (solid curves). The dependence on $V_{\rm gate}$ at a fixed value of $W/L$ is shown in Fig.\ \ref{SigmaGate}.

Fig.\ \ref{Sigma0} also contains results for a boundary condition corresponding to an abrupt edge (dashed curves). We considered a ``metallic armchair'' edge, in which the carbon lattice contains a multiple of three hexagons in the transverse direction, terminated at $y=0$ and $y=W$ by a horizontal bond. This edge mixes the valleys, so we need to consider a four-component wave function $\Psi=(\Psi_{1},\Psi_{2},\Psi'_{1},\Psi'_{2})$. The first two components satisfy the Dirac equation (\ref{Dirac}), without the mass term, and the second two components satisfy the same equation with $p_{y}\rightarrow -p_{y}$. The boundary condition is \cite{Bre06}
\begin{equation}
\Psi_{1}=\Psi'_{1},\;\;\Psi_{2}=\Psi'_{2},\;\;{\rm at}\;\;y=0,W.\label{armchairbc}
\end{equation}

The valley degeneracy is broken for the lowest mode ($n=0$), which is nondegenerate, while all higher modes ($n=1,2,\ldots$) retain the twofold valley degeneracy (over and above the twofold spin degeneracy, common to all modes). For $V_{\rm gate}=0$ and $N\rightarrow\infty$ the transmission probabilities are given by
\begin{equation}
T_{n}=\frac{1}{\cosh^2( \pi nL/W)},\;\;n=0,1,2,\ldots\,.\label{tn_armchair}
\end{equation}
The essential difference with the result (\ref{tn_hf}) for the smooth edge is due to the absence of the $1/2$ offset in the quantization condition of the transverse momentum. The different boundary condition changes the strip from insulating to metallic in the limit $W/L\rightarrow 0$ \cite{note2}, but has no effect in the opposite limit $W/L\rightarrow \infty$, cf.\ Fig.\ \ref{Sigma0}.

To test the analytical results from the continuum description of graphene, we have also carried out numerical simulations using the tight-binding model with nearest-neighbor hopping on a honeycomb lattice with metallic armchair edges. We took some 2000 lattice sites for the graphene strip, coupling it to semi-infinite leads at the two ends. The valley degeneracy of modes $n=1,2,\ldots$ is now only approximate, but the relative magnitude of the mode splitting vanishes $\propto a/W$ as the width becomes large compared to the lattice spacing $a$ \cite{Wak99}. The numerical results, included in Fig.\ \ref{Sigma0}, are in excellent agreement with the analytical prediction.

Fig.\ \ref {SigmaGate} shows that the minimum in the conductivity at the Dirac point is associated with a maximum in the Fano factor. The limiting behavior at the Dirac point for a short and wide strip is
\begin{eqnarray}
\sigma\rightarrow g_{0}/\pi,\;\; F\rightarrow 1/3,\;\; {\rm for}\;\;W/L\rightarrow\infty.\label{limits}
\end{eqnarray}
Note that these limits are already reached for moderate aspect ratios $W/L \gtrsim 4$. We derived this limiting behavior for two types of boundary conditions (smooth edge and metallic armchair edge), but we are confident that the result is universal, in the sense that it holds for the most general boundary condition at the edges of the graphene strip (as classified in Ref.\ \cite{McC04}).

The result (\ref{limits}) for the minimal conductivity agrees with other calculations \cite{Lud94,Zie98,Per05}, which start from an unbounded disordered system and then take the limit of infinite mean free path $l$. There is no geometry dependence if the limits are taken in that order. In the ballistic bounded system considered here, a geometry dependence persists in the thermodynamic limit \cite{Kat05}. Existing experiments \cite{Nov05,Zha05} are quasi-ballistic, with $l\simeq W<L$, finding $\sigma\approx g_{0}$. This would be consistent with Fig.\ \ref{Sigma0}a if the effective length in the experiment is set by $l$ rather than $L$. Residual disorder may also explain why the oscillations in Fig.\ \ref{SigmaGate}a, associated with the appearance of propagating modes, are not observed in the experimental gate voltage dependence (compare figure 1 of Ref.\ \cite{Nov05}).

The limit $F=1/3$ for the Fano factor is smaller than the value $F=1$ expected for a Poisson process. The same $1/3$ value appears in a disordered metal \cite{Bee92,Nag92}, where it is a consequence of classical diffusive dynamics. This correspondence is remarkable, since in our ideal graphene strip the classical dynamics is ballistic. The relativistic quantum dynamics of confined Dirac fermions is known to exhibit a jittering motion called ``Zitterbewegung'' \cite{Sch30}, originating from the interference of states with positive and negative energy \cite{Kat05,Zaw05}. Our calculation implies that this relativistic quantum dynamics produces the same shot noise as classical diffusion.

In conclusion, we predict that electrical conduction through an ideal graphene strip is associated with time-dependent current fluctuations --- at zero temperature and without any impurities or lattice defects. The electrical noise is largest, relative to the mean current, when the Fermi energy is adjusted such that electrons and holes are degenerate. At this Dirac point the Fano factor (ratio of noise power and mean current) takes on the universal value $1/3$ for short and wide strips. Observation of this quantum-limited shot noise would be a unique demonstration of electrical noise produced by relativistic quantum dynamics.

We thank H. A. Fertig and A. Morpurgo for interesting discussions. This research
was supported by the Dutch Science Foundation NWO/FOM and by the European Community's Marie Curie Research Training Network (contract MRTN-CT-2003-504574, Fundamentals of Nanoelectronics). AR acknowledges a Foreign Postdoc Fellowship from the Polish Science Foundation.

\appendix
\section{Calculation of the transmission probability}

\subsection{Mode matching}

Plane wave solutions of the Dirac equation
\begin{equation}
\frac{\hbar v}{i}\begin{pmatrix}
\sigma_{x}\partial_{x}+\sigma_{y}\partial_{y}&0\\
0&\sigma_{x}\partial_{x}-\sigma_{y}\partial_{y}
\end{pmatrix}\Psi+eV\Psi=\varepsilon\Psi
\end{equation}
take the form
\begin{eqnarray}
\Psi_{n,k}(\bm{r})&=&\chi_{n,k}(y)e^{ikx},\label{Psibasisdef}\\
\chi_{n,k}(y)&=&a_{n}
\begin{pmatrix}1\\ z_{n,k}\\ 0\\ 0\end{pmatrix}
e^{iq_{n}y}+
a'_{n}
\begin{pmatrix}0\\ 0\\ z_{n,k}\\ 1\end{pmatrix}
e^{iq_{n}y}\nonumber\\
&&\mbox{}+
b_{n}
\begin{pmatrix}z_{n,k}\\ 1\\ 0\\ 0\end{pmatrix}
e^{-iq_{n}y}+
b'_{n}
\begin{pmatrix}0\\ 0\\ 1\\ z_{n,k}\end{pmatrix}
e^{-iq_{n}y}.\nonumber\\
&&
\end{eqnarray}
The complex number $z_{n,k}$ is defined by
\begin{equation}
z_{n,k}=\pm\frac{k+iq_{n}}{\sqrt{k^{2}+q_{n}^{2}}},
\end{equation}
where the $+$ sign applies to the conduction band and the $-$ sign to the valence band. For later use we note the identity
\begin{equation}
z_{n,k}z_{n,-k}=-1.\label{zidentity}
\end{equation}

The energy of the state in the leads ($x<0$ and $x>L$, where $V=V_{\infty}$) is given by
\begin{equation}
\varepsilon=eV_{\infty}\pm\hbar v\sqrt{k^{2}+q_{n}^{2}}.
\end{equation}
In the strip ($0<x<L$, where $V=V_{\mathrm{gate}}$) the wave vector $k$ is replaced by $\tilde{k}$, satisfying
\begin{equation}
\varepsilon=eV_{\mathrm{gate}}\pm\hbar v\sqrt{\tilde{k}^{2}+q_{n}^{2}}.\label{ktildedef}
\end{equation}
(The $\pm$ sign again refers to conduction and valence bands.) In the leads we have only propagating modes (real $k$), while the modes in the strip may be propagating (real $\tilde{k}$) or evanescent (imaginary $\tilde{k}$). The sign of $\tilde{k}$ is not fixed by Eq.\ (\ref{ktildedef}) and may be chosen arbitrarily.

The transverse wave vector $q_{n}$ as well as the coefficients $a_{n},a'_{n},b_{n},b'_{n}$ of the $n$-th mode are determined (up to a normalization constant) by the boundary conditions at $y=0$ and $y=W$. We consider a class of boundary conditions for which the resulting parameters are independent of the longitudinal wave vectors $k,\tilde{k}$. Because $q_{n}$ is then the same in the strip and in the leads, the mode matching at the lead--strip interface does not mix the modes and can be done analytically.

A scattering state $\Psi$ incident from the left at energy $\varepsilon$ in the $n$-th mode is constructed from the states (\ref{Psibasisdef}),
\begin{eqnarray}
&&\Psi=\left\{\begin{array}{ll}
\Phi_{L}&\mathrm{if}\;\;x<0,\\
\tilde{\Phi}&\mathrm{if}\;\;0<x<L,\\
\Phi_{R}&\mathrm{if}\;\;x>L,
\end{array}\right.\\
&&\Phi_{L}=\chi_{n,k}e^{ikx}+r_{n}\chi_{n,-k}e^{-ikx},\\
&&\Phi_{R}=t_{n}\chi_{n,k}e^{ik(x-L)},\\
&&\tilde{\Phi}=\alpha_{n}\chi_{n,\tilde{k}}e^{i\tilde{k}x}+\beta_{n}\chi_{n,-\tilde{k}}e^{-i\tilde{k}x}.
\end{eqnarray}
The sign of the (real) wave vector $k$ is chosen positive for the conduction band and negative for the valence band.

The reflection and transmission amplitudes $r_{n},t_{n}$ and the two coefficients $\alpha_{n},\beta_{n}$ are determined by requiring continuity of $\Psi$ at $x=0$ and $x=L$. There are four independent equations,
\begin{eqnarray}
&&\begin{pmatrix}1\\ z_{n,k}\end{pmatrix}+
r_{n}\begin{pmatrix}1\\ z_{n,-k}\end{pmatrix}=
\alpha_{n}\begin{pmatrix}1\\ z_{n,\tilde{k}}\end{pmatrix}+
\beta_{n}\begin{pmatrix}1\\ z_{n,-\tilde{k}}\end{pmatrix},\nonumber\\
&&\\
&&t_{n}\begin{pmatrix}1\\ z_{n,k}\end{pmatrix}=
\alpha_{n}\begin{pmatrix}1\\ z_{n,\tilde{k}}\end{pmatrix}e^{i\tilde{k}L}+
\beta_{n}\begin{pmatrix}1\\ z_{n,-\tilde{k}}\end{pmatrix}e^{-i\tilde{k}L}.\nonumber\\
&&
\end{eqnarray}
Notice that the coefficients $a_{n},a'_{n},b_{n},b'_{n}$ have dropped out. We need the solution for the transmission amplitude,
\begin{equation}
t_{n}=\frac{(1+z_{n,k}^{2})(1+z_{n,\tilde{k}}^{2})}{e^{i\tilde{k}L}(z_{n,k}-z_{n,\tilde{k}})^{2}+e^{-i\tilde{k}L}(1+z_{n,k}z_{n,\tilde{k}})^{2}},
\end{equation}
where we have used the identity (\ref{zidentity}).
The transmission probability $T_{n}$ of the $n$-the mode at the Fermi level is obtained from $T_{n}=|t_{n}|^{2}$ at $\varepsilon=0$.

We are interested in the limit $|V_{\infty}|\rightarrow\infty$ of an infinite number of propagating modes in the leads. Then $z_{n,k}\rightarrow 1$, hence
\begin{eqnarray}
T_{n}&=&\left|\frac{2+2z_{n,\tilde{k}}^{2}}{e^{i\tilde{k}L}(1-z_{n,\tilde{k}})^{2}+e^{-i\tilde{k}L}(1+z_{n,\tilde{k}})^{2}}\right|^{2}\nonumber\\
&=&\left|\frac{\tilde{k}}{\tilde{k}\cos(\tilde{k}L)+i(eV_{\mathrm{gate}}/\hbar v)\sin(\tilde{k}L)}\right|^{2},\label{Tn_qn_result}
\end{eqnarray}
with $\tilde{k}^{2}=(eV_{\mathrm{gate}}/\hbar v)^{2}-q_{n}^{2}$. This result corresponds to Eq.\ (\ref{Tna}) in the main text.

\subsection{Boundary conditions}

It remains to calculate the transverse wave vector $q_{n}$, for specific boundary conditions at $y=0$ and $y=W$. Substitution into Eq.\ (\ref{Tn_qn_result}) then completes the calculation of the transmission probabilities.

For the infinite mass boundary condition,
\begin{eqnarray}
&&\Psi|_{y=0}=\begin{pmatrix}\sigma_{x}&0\\ 0&-\sigma_{x}\end{pmatrix}\Psi|_{y=0},\\
&&\Psi|_{y=W}=\begin{pmatrix}-\sigma_{x}&0\\ 0&\sigma_{x}\end{pmatrix}\Psi|_{y=W},
\end{eqnarray}
one finds
\begin{equation}
q_{n}=(n+\tfrac{1}{2})\frac{\pi}{W},\;\;\frac{a_{n}}{b_{n}}=-\frac{a'_{n}}{b'_{n}}=1.
\end{equation}
Each mode $n=0,1,2,\ldots$ has a twofold valley degeneracy.

For the metallic armchair edge,
\begin{eqnarray}
&&\Psi|_{y=0}=\begin{pmatrix}0&1\\ 1&0\end{pmatrix}\Psi|_{y=0},\\
&&\Psi|_{y=W}=\begin{pmatrix}0&e^{i\phi}\\ e^{-i\phi}&0\end{pmatrix}\Psi|_{y=W},\label{armchairbc2}
\end{eqnarray}
with $\phi=0$, one finds
\begin{equation}
q_{n}=\frac{n\pi}{W},\;\;\frac{a_{n}}{b'_{n}}=\frac{a'_{n}}{b_{n}}=1.
\end{equation}
The modes with $n=1,2,3\ldots$ have a twofold degeneracy (because the choices $a_{n}=0$ and $a'_{n}=0$ produce two independent states). The lowest mode $n=0$, however, is nondegenerate (because $|z_{0,k}|=1$, so the states produced by $a_{n}=0$ and $a'_{n}=0$ are the same).

A semiconducting armchair edge (corresponding to a strip width which is not an integer multiple of three unit cells) has $\phi=\pm 2\pi/3$ in Eq.\ (\ref{armchairbc2}). Then all modes are nondegenerate, consisting of one sequence with $q_{n}=(n-\phi/2\pi)\pi/W$, $a_{n}=b'_{n}=0$, $a'_{n}=b_{n}$ and one sequence with $q_{n}=(n+\phi/2\pi)\pi/W$, $a'_{n}=b_{n}=0$, $a_{n}=b'_{n}$.

The results for the conductance and Fano factor at the Dirac point, including all degeneracies, may be summarized by
\begin{eqnarray}
&&G=\frac{2e^{2}}{h}\sum_{n=-\infty}^{\infty} T_{n} ,\;\; F=\frac{\sum\limits_{n=-\infty}^{\infty}T_{n}(1-T_{n})}{\sum\limits_{n=-\infty}^{\infty}T_{n}},\label{GFgeneral}\\
&&T_{n}=\frac{1}{\cosh^{2}[(n+\alpha)\pi L/W]},\label{Tngeneral}
\end{eqnarray}
with $\alpha=1/2$ for the infinite mass boundary condition, $\alpha=0$ for the metallic armchair edge, and $\alpha=1/3$ for the semiconducting armchair edge.

The general formula (\ref{Tngeneral}) holds under the assumption that longitudinal and transverse momenta are not coupled by the boundary condition. The zigzag edge is an example of a boundary which does couple $k$ and $q_{n}$. In that case the jump in $k$ at $x=0$ and $x=L$ introduces a mode mixing which complicates the analytical solution. We do not expect any differences to appear in the limit (\ref{limits}) for $W/L\rightarrow\infty$.


\begin{thebibliography}{99}
\bibitem{Nov05} K. S. Novoselov, A. K. Geim, S. V. Morozov, D. Jiang,
M. I. Katsnelson, I. V. Grigorieva, S. V. Dubonos, and A. A. Firsov,
Nature {\bf 438}, 197 (2005).
\bibitem{Zha05} Y. Zhang, Y.-W. Tan, H. L. Stormer, and P. Kim,
Nature {\bf 438}, 201 (2005).
\bibitem{Lud94} A. W. W. Ludwig, M. P. A. Fisher, R. Shankar, and G. Grinstein,
Phys.\ Rev.\ B {\bf 50}, 7526 (1994).
\bibitem{Zie98} K. Ziegler,
Phys.\ Rev.\ Lett.\ {\bf 80}, 3113 (1998).
\bibitem{Per05} N. M. R. Peres, F. Guinea, and A. H. Castro Neto,
Phys.\ Rev.\ B {\bf 73}, 125411 (2006).
\bibitem{Kat05} M. I. Katsnelson,
cond-mat/0512337.
\bibitem{Bee92} C. W. J. Beenakker and M. B\"{u}ttiker,
Phys.\ Rev.\ B {\bf 46}, 1889 (1992).
\bibitem{Nag92} K. E. Nagaev,
Phys.\ Lett.\ A {\bf 169}, 103 (1992).
\bibitem{Roc02}
P.-E. Roche, M. Kociak, S. Gu{\'e}ron, A. Kasumov, B. Reulet, and
H. Bouchiat, 
Eur.\ Phys.\ J. B {\bf 28}, 217 (2002).
\bibitem{Ona06}
E. Onac, F. Balestro, B. Trauzettel, C. F. J. Lodewijk, and L. P. Kouwenhoven,
Phys.\ Rev.\ Lett.\ {\bf 96}, 026803 (2006).
\bibitem{Per06} N. M. R. Peres, A. H. Castro Neto, and F. Guinea,
cond-mat/0512476. 
\bibitem{Bre06} L. Brey and H. A. Fertig, 
cond-mat/0603107.
\bibitem{Ber87} M. V. Berry and R. J. Mondragon, 
Proc.\ R. Soc.\ Lond.\ A {\bf 412}, 53 (1987).
\bibitem{note1} The transmission probability of nonrelativistic electrons through a rectangular barrier at zero energy (relative to the top of the barrier) is given by $T_{n}=[\cosh^{2}Lq_{n}+(k_{\infty}^{2}/2q_{n}^{2}-1)^{2}(k_{\infty}^{2}/q_{n}^{2}-1)^{-1}\sinh^{2}Lq_{n}]^{-1}$, with $q_{n}=n\pi/W$ ($n=1,2,\ldots$). In contrast to the relativistic result (\ref{tn_hf}), it vanishes $\propto (q_{n}/k_{\infty})^{2}$ in the limit $k_{\infty}\rightarrow \infty$.
\bibitem{note2} For $W/L\rightarrow 0$ we find $G\rightarrow g_{0}/2$ in the case of a metallic armchair edge. The factor $1/2$, due to the absence of valley degeneracy for the lowest mode, is missing from Ref.\ \cite{Per06}.
\bibitem{McC04} E. McCann and V. I. Fal'ko,
J. Phys.\ Condens.\ Matter {\bf 16}, 2371 (2004).
\bibitem{Wak99}
The magnitude of the mode splitting follows from the
formulas given in the appendix of
K. Wakabayashi, M. Fujita, H. Ajiki, and M. Sigrist,
Phys.\ Rev.\ B {\bf 59}, 8271 (1999).
\bibitem{Sch30} E. Schr\"{o}dinger,
Sitzber.\ Preu{\ss}. Akad.\ Wiss.\ {\bf 24}, 418 (1930).
\bibitem{Zaw05} W. Zawadzki,
cond-mat/0510184.
\end{thebibliography}
\end{document}